\documentclass[
showpacs,
amsmath,amssymb,
aps,
twocolumn,
prb,
floatfix,
nofootinbib
]{revtex4-2}
\usepackage{amsmath}
\usepackage[utf8]{inputenc}
\usepackage[T1]{fontenc}
\usepackage{braket}
\usepackage{subcaption}
\usepackage{float}

\usepackage{lipsum}
\usepackage{graphicx}
\usepackage[switch]{lineno}
\usepackage{hyperref}
\usepackage{xcolor}
\usepackage{url}
\usepackage{natbib}
\usepackage{placeins}

\begin{document}

\title{ Intrinsic space-time crystalline order in hybrid Josephson junction}

\author{M. Nashaat$^{1,2,*}$, J. Tekić$^{3,\dagger}$, and Yu. M. Shukrinov$^{1,4,\ddagger}$}
\affiliation{$^1$\mbox{BLTP, JINR, Dubna, Moscow region, 141980, Russia}\\
$^2$ \mbox{Department of Physics, Faculty of Science, Cairo University, 12613, Giza, Egypt}\\
  \vbox{$^3$ "Vinča" Institute of Nuclear Sciences, Laboratory for Theoretical and Condensed Matter Physics - 020, \\ University of Belgrade, PO Box 522, 11001 Belgrade, Republic of Serbia} \\
$^4$ \mbox{Dubna State University, Dubna, Russia}\\
$^{*}$majed@sci.cu.edu.eg; $^{\dagger}$jasminat@vin.bg.ac.rs ;$^{\ddagger}$shukrinv@theor.jinr.ru
}


\begin{abstract}
We demonstrate the emergence of an intrinsic space-time crystalline order in a long ferromagnetic $\varphi_0$ Josephson junction on a topological insulator without any external periodic modulation.
The presence of the exchange and Dzyaloshinskii–Moriya interactions in a ferromagnetic layer with broken inversion symmetry internally modulates the critical current due to the coupling between the magnetic moment and the Josephson phase. 
This breaks the time translation symmetry, leading to the appearance of the space-time crystalline pattern in the spatiotemporal dependence of the in-plane current, which oscillates with almost twice the ferromagnetic resonance frequency.
In the limit where the critical current is not modulated internally, the space-time crystalline order does not occur. 
In this case, only when an external parametric modulation is applied, the system exhibits a typical classical discrete space-time crystalline order that oscillates at half of the modulation frequency.
Considering the still-pending problem of experimental detection, we demonstrate that a recently developed magnetometry device, which visualizes the supercurrent flow at the nanoscale, can be used to detect space-time crystalline patterns in hybrid Josephson junctions experimentally.

\end{abstract}

\flushbottom
\maketitle


\section{Introduction}
\label{Intr}

Since their first proposal by Frank Wilczek~\cite{wilczek2012quantum,shapere2012classical}, time crystals (TC) have been causing a great stir in the scientific community. 
After the initial controversies on their existence~\cite{watanabe2015absence,bruno2013impossibility}, they are now considered as a nonequilibrium phase of matter that can operate in the time dimension in a manner akin to the standard crystals in the space dimension.
The defining characteristic of these systems is the breaking of discrete or continuous time translation symmetry, leading to a self-sustained and robust time-periodic order. 
Since the mid-2010s, time crystals have been conceptualized in both quantum and classical systems~\cite{hannaford2022decade,zaletel2023colloquium}, followed by a huge effort invested in their experimental realization~\cite{zhang2017observation,choi2017observation,abanin2019colloquium,kessler2021observation,kongkhambut2022observation,liu2023photonic,chen2023realization,wang2025expanding}.
Recently, a robust continuous time crystal was observed in an electron–nuclear spin system~\cite{greilich2024robust}. 

Despite a significant number of works produced, most of these studies explore quantum time crystals, while a relatively smaller number have focused on classical systems, especially on time crystals in condensed-matter physics~\cite{kleiner2021space,yao2020classical,hannaford2022decade,guo2020condensed}. 
Recently, a discrete space-time crystal was proposed in the high-$T_c$ superconductor with intrinsic Josephson junctions, where under the periodic parametric modulation, the Josephson current developed half-harmonic oscillations in time and broke continuous translation symmetry in space~\cite{kleiner2021space}.

Another question still troubling scientists is related to the original idea that TC should represent self-generated and self-sustained motion without any external input~\cite{wilczek2012quantum}.
As the "no-go" theorem proved it impossible~\cite{watanabe2015absence,bruno2013impossibility}, for TC to survive those controversies~\cite{watanabe2015absence}, this idea had to be abandoned, and the realization of TC has relied on imposing external periodic modulations~\cite{kleiner2021space} or modulation through cavity feedback~\cite{kongkhambut2022observation,kessler2019emergent} on the system. 
All the achievements made so far, along with the still existing challenges of creating a TC without external influence, only further fuel the desire of scientists to get as close as possible to the impossible Wilczek idea~\cite{chen2023realization}.
Recent progress was made in Ref.~\cite{chen2023realization}, showing that an ensemble of pumped erbium ions could give rise to an inherent time-crystalline phase.

In this work, we investigate the occurrence of space-time crystalline (STC) order without any external periodic input in a hybrid Josephson junction with a ferromagnetic interface. 
These junctions offer a unique paradigm, where superconductivity and magnetism can coexist and interact~\cite{shukrinov2022anomalous,amundsen2024colloquium}.
The presence of a ferromagnetic layer introduces the magnetic moment into the dynamics of a system, resulting in the appearance of a ferromagnetic resonance (FMR), which occurs when the Josephson frequency becomes close to that of the FMR one~\cite{shukrinov2019ferromagnetic,shukrinov2022anomalous}. 
Unlike in conventional Josephson junctions, here the dynamics of the ferromagnetic moment can influence the Josephson current and vice versa.

\begin{figure}[h!]
	\centering
	\includegraphics[width=0.8\linewidth]{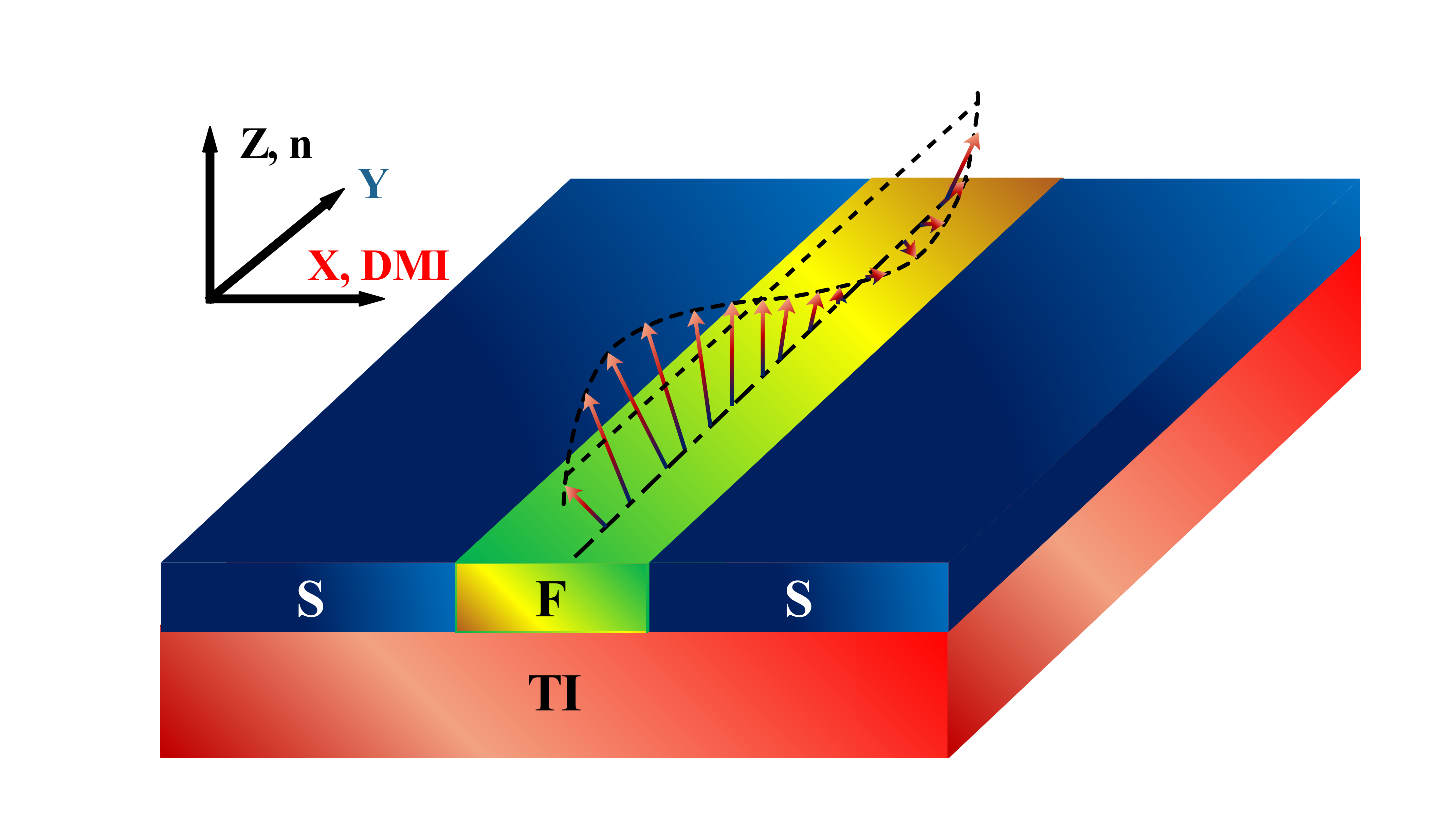}
	\caption{Proposed long $\varphi _0$ Josephson junction deposited on top of the three-dimensional topological insulator. The arrows show the precession of magnetization in the F-layer.}
	\label{Fig1}
\end{figure}

We consider a long superconductor-ferromagnetic-superconductor Josephson junction on a topological insulator (SFS-TI JJ)~\cite{nashaat2019electrical}, presented in Fig.~\ref{Fig1},
with a noncentrosymmetric ferromagnetic layer and broken time-reversal symmetry, where the precession of the magnetic moment is modified by the Dzyaloshinskii–Moriya interaction (DMI)~\cite{hong2017magnetic}.
The Rashba-type spin-orbit coupling in the ferromagnetic layer leads to an additional phase shift $\varphi_0$ in the current-phase relation (CPR) proportional to the magnetic moment in the barrier and the strength of the spin-orbit coupling~\cite{buzdin2008direct,konschelle09, guarcello,abdelmoneim2022locking,guarcello2023switching}.
As a result, the CPR takes the form $I = I_{c} \sin(\varphi-\varphi_{0})$, where $ I_{c}$ is the critical current and $\varphi$ is the superconducting phase difference.
This type of SFS JJ is simply called $\varphi_{0}$ JJ.
If the $\varphi_{0}$ JJ is placed on top of a topological insulator (TI), then the critical current strongly depends on the in-plane magnetization component along the current direction~\cite{nashaat2019electrical}. 
As we will show, this dependence, i.e., the internal modification of the superconducting critical current by the magnetic moment, can pave the way to the creation of space-time crystalline order in this junction.

The paper is organized as follows:
in Sec.~\ref{Mod} we introduce the model of a long $\varphi _0$ Josephson junction placed on a topological insulator.
In Sec.~\ref{SFS_TI JJ}, we consider the case when the Josephson current is internally modulated by the magnetic moment, where in Sec.~\ref{STC} we investigate the spatiotemporal dependence of the in-plane current. 
To confirm the STC order and find the frequency at which the in-plane current oscillates, we also analyze the averaged current-current correlation function and perform the fast Fourier transform analysis (FFT).
The influence of system parameters and the robustness of the STC order are also examined.
In Sec.~\ref{M}, we investigate the magnetic subsystem, which does not display STC order.
In Sec.~\ref{SFS JJ}, to show that the modulation of Josephson critical current is the key requirement for the occurrence of STC order, we consider the limit when the critical current is not modulated by the magnetic moment.
This results in the disappearance of the STC pattern, and the STC order can be realized only if an external periodic parametric modulation is applied.
In Sec.~\ref{Dis}, we discuss the results presented and provide insight into the physics behind them.
In Sec.~\ref{Exp}, we propose a method for the experimental detection of STC order in the in-plane current.
Finally, Sec.~\ref{Con} concludes the paper.

\section{Model}
\label{Mod}

We consider a long $\varphi _0$ JJ on TI where the easy axis of magnetization in the F-layer is in the $y$ direction and the unit vector $\textbf{n}$ is normal to the surface of TI (see Fig.~\ref{Fig1}). 
In this particular geometry $\varphi _0=rm_y$, which leads to the current phase relation given by 
$j_{c}(m_{x}) \sin(\varphi-rm_{y})$~\cite{nashaat2019electrical}, where $r$ characterizes the strength of spin-orbit coupling, and $m_{i},(i=x,y)$ is the normalized magnetization components to the saturation value $M_{s}$. 
The DMI vector is perpendicular to both the inversion asymmetry direction and the vector between $\textbf{m}_{i}$ and $\textbf{m}_{j}$. Thus, in our case, the DMI vector is along the $x$-axis.

\begin{figure}[h!]
	\centering
        \includegraphics[width=0.8\linewidth]{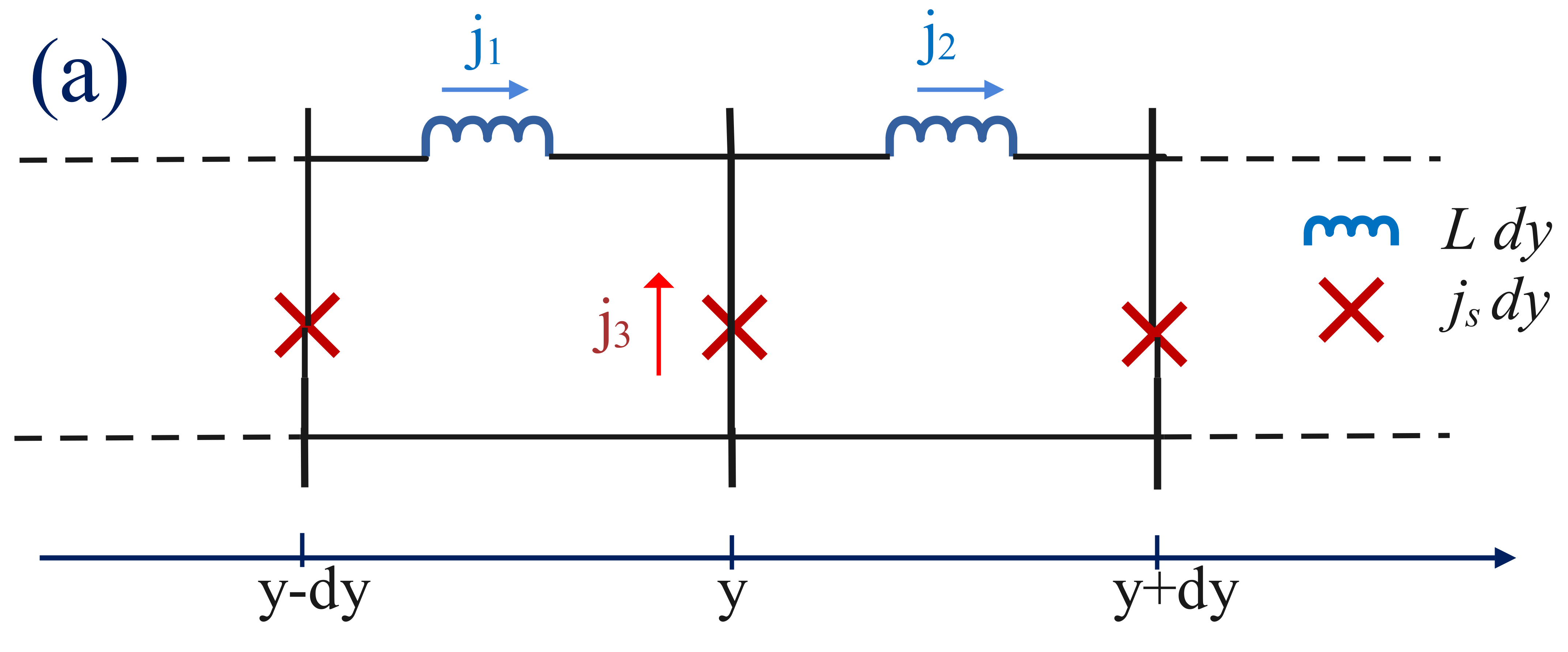}
	\includegraphics[width=0.8\linewidth]{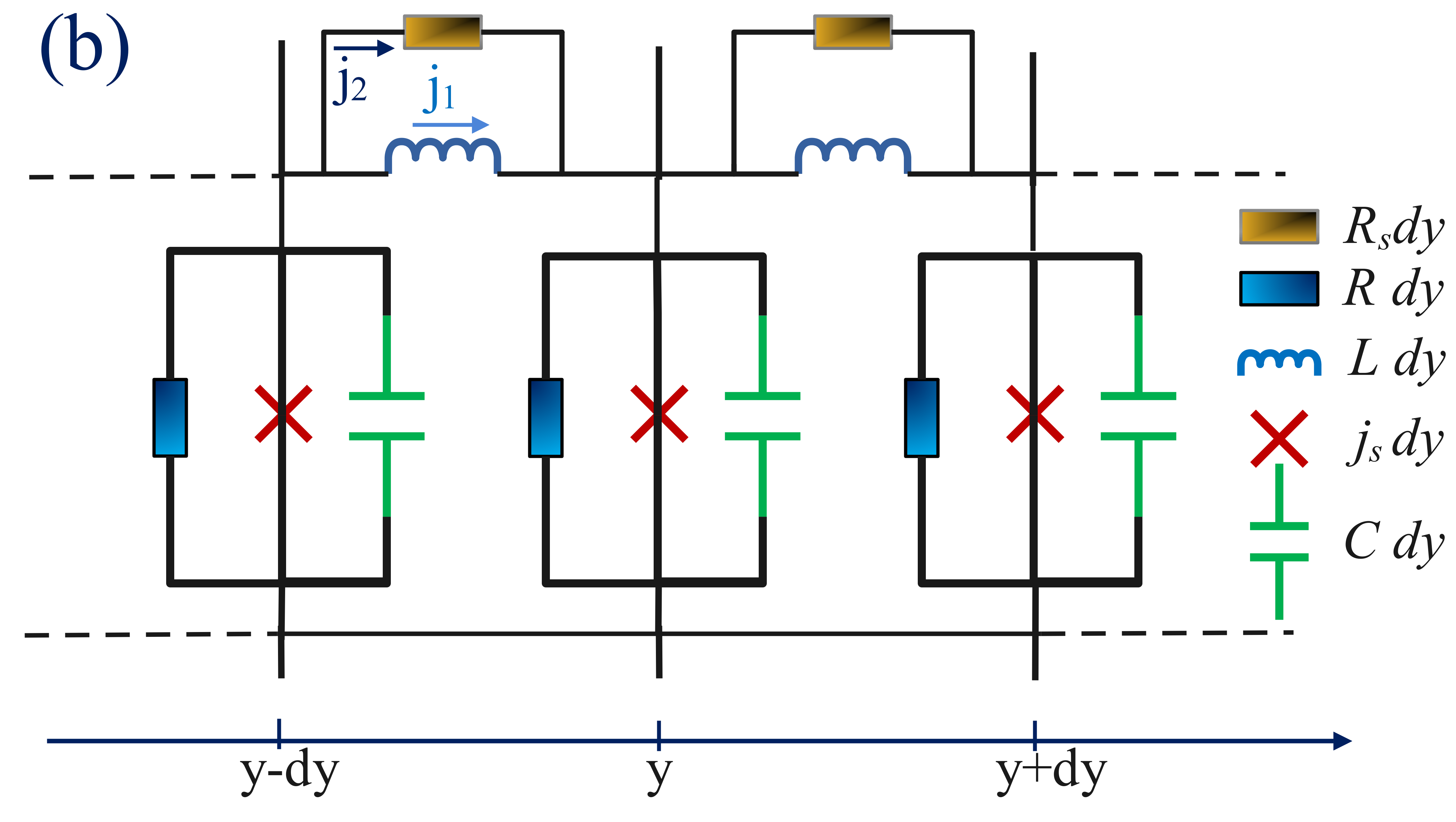}
	\caption{(a) Equivalent circuit for the long $\varphi _0$ Josephson junction. Each discrete element in the model is characterized by a Josephson current branch $j_{s}$. The discrete elements are inductively coupled, with currents $j_{1}$ and $j_{2}$ passing through the inductor branches. (b) The same as in (a) but with the quasiparticle current branch characterized by $R$, and the displacement current characterized by $C$. The lossy current is represented by the branch with resistance $R_{s}$.}
	\label{Fig2}
\end{figure}

To describe the dynamics of a long $\varphi_0$ JJ, we modified the sine-Gordon equation~\cite{visser2002modelling,rahmonov2014parametric,golovchanskiy2017ferromagnetic} using a discrete system method~\cite{visser2002modelling}. 
In this approach, the relevant physical properties of the junction, such as resistance, capacitance, etc., are assumed to be constant over some interval of the coordinate ($\Delta y$). 
Then, the equations for the long junction are obtained by taking the limit $\Delta y \rightarrow 0$. 
Accordingly, a discrete circuit with a specific impedance shown in Fig.~\ref{Fig2} is used to model the resistance, capacitance, and inductance effects.

We consider a one-dimensional (1D) array of SFS JJs with inductive coupling, and \textit{no external bias or periodic input.}
The sine-Gordon equation at low temperatures for ideal superconductors (no quasiparticle currents) can be obtained by considering the circuit shown in Fig.~\ref{Fig2}(a) with the corresponding currents:
\begin{eqnarray}
     j_{1}&=&\frac{\Phi_{0}}{2\pi} \frac{ \varphi(y)-\varphi(y-dy) }{Ldy}, \nonumber \\ 
     j_{2}&=&\frac{\Phi_{0}}{2\pi} \frac{ \varphi(y+dy)-\varphi(y) }{Ldy}, \nonumber \\
     j_{3}&=&j_{c} (m_{x}) \sin(\varphi-rm_{y}) dy,
\end{eqnarray}
where $\Phi_{0}$ is the flux quantum, $L$ is the inductance, $j_c$ is the critical current density and $m_{x,y}$ are the magnetic components. 
At the nodes, the Kirchhoff law is satisfied, such that
$j_{2}-j_{1}+j_{3}=0$, which leads us to the sine-Gordon equation for Josephson phase:
\begin{eqnarray}
    \frac{\Phi_{0}}{2\pi L} \frac{d^{2}\varphi}{dy^{2}}= j_{c}(m_{x})\sin(\varphi-rm_{y}).
    \label{dc}
\end{eqnarray}
This equation also describes the "dc" Josephson effect ($\partial \varphi/\partial t=0$).

At higher temperatures, the superconductor will contain some quasiparticles, causing lossy currents. 
It is also possible that a pair of electrons can separate if the superconductor leads have a significant potential difference, and one of the separated electrons tunnels to the other superconductor. 
Consequently, a nonzero resistance $R_s$ arises. 
To simplify this case, we consider that the inductive coupling between the junctions is only due to the Josephson phase. 
This means that we neglect the magnetic field due to the quasiparticle current resulting from the magnetic precession. 
In this case, we consider the circuit shown in Fig.~\ref{Fig2}(b) described by the following equations:
\begin{eqnarray}
\frac{\partial V}{\partial y}&=&-j_{2} R_{s} =-L\frac{\partial j_{1}}{\partial t}, \\
\frac{\partial (j_{1}+j_{2})}{\partial y} &=& -C \frac{\partial V}{\partial t} - \frac{V}{R}  \nonumber \\&-& j_{c} (m_{x}) \sin(\varphi-r m_{y}) +\frac{r \Phi_{0}}{2 \pi R}\frac{\partial m_{y}}{\partial t}.
\label{ac-equ}
\end{eqnarray}
After using the Josephson relation , $\partial\varphi/\partial t= (2\pi/\Phi_{0}) V$, we can write $\frac{\partial j_{1}}{\partial t}$ and $j_{2}$ as 
\begin{eqnarray}
    \frac{\partial j_{1}}{\partial t} =-\frac{1}{L} \frac{\partial V}{\partial y} =-\frac{\Phi_{0}}{2\pi L} \frac{\partial ^{2} \varphi}{\partial y \partial t}, \nonumber \\
    j_{2}= -\frac{1}{R_{s}} \frac{\partial V}{\partial y}=-\frac{\Phi_{0}}{2\pi R_{s}} \frac{\partial ^{2} \varphi}{\partial y \partial t}. 
    \label{j1j2}
\end{eqnarray}
By inserting Eq.~(\ref{j1j2}) into (\ref{ac-equ}), we obtain the sine-Gordon equation, which in the dimensionless form is given as
\begin{eqnarray}
&&\alpha \frac{\partial^{2}}{\partial y^{2}} \frac{\partial \varphi}{\partial t} +  \frac{\partial^{2} \varphi}{\partial y^{2}} - \frac{\partial^{2} \varphi}{\partial t^{2}}  -\beta \bigg( \frac{\partial \varphi}{\partial t} - r \frac{\partial  m_{y}}{\partial t}\bigg)\nonumber\\&-& J_{c}(m_{x})\sin (\varphi-r m_{y})+J_{noise}  =0, 
\label{sine-Gor_eq}
\end{eqnarray}
with the following boundary conditions:
\begin{eqnarray}
\frac{\partial \varphi}{\partial y}|_{y=0(L)} &+&\alpha\frac{\partial^{2} \varphi}{\partial y\partial t}|_{y=0(L)}=0,
\end{eqnarray}
where $\beta $ is the dissipation coefficient (here it is assumed that $\beta\ll 1$). 
The coordinate $y$ is normalized to Josephson penetration depth $\lambda_{J}=\sqrt{\Phi_{0}/(2\pi \mu J_{c} d_{J})}$, where $\mu$ is the permeability, $d_{J} = 2\lambda_{L}+t_{b}$ is the thickness of the magnetic layer, $t_{b}$ is the thickness of the tunneling barrier, and $\lambda_{L}$ is the London penetration depth.
The time $t$ is normalized in units of $\omega_{J}^{-1}$, where $\omega_{J}$ is the Josephson plasma frequency, which can be in the range of gigahertz~\cite{golovchanskiy2017ferromagnetic}. 
The noise current $J_{noise}$ is added to keep the system out of equilibrium and normalized to $J_{c}$ when $m_{x}=0$.

We need to stress the following points.
First, we neglect the surface loss ($\alpha=0$).
Second, unlike models that assign a separate RLC branch to the ferromagnet~\cite{petkovic2009direct}, our equivalent circuit for the proposed junction integrates its effect directly into the critical current and Josephson current-phase relation.
Third, the displacement current is proportional to the first derivative of the voltage, which is determined by the phase difference and does not depend on $\varphi_{0}$. 
From this point of view, we do not include the second derivative of $\varphi_{0}$ in our model~\cite{janalizadeh2022nonlinear,guarcello2023switching}.
Fourth, the magnetization dynamics plays the role of a driving force, and the first-order derivative of $\varphi_{0}=rm_{y}$ is a source of quasiparticle current in the JJ. 

The modified sine-Gordon equation Eq.~(\ref{sine-Gor_eq}) for the Josephson phase difference $\varphi$ across the long JJ is coupled with the Landau-Lifshitz-Gilbert equation for the magnetization dynamics~\cite{miltat2007numerical}:
\begin{eqnarray}
\frac{d\mathbf{m}}{dt}=-\frac{\Omega_{F}}{(1+\alpha_{g}^{2})}\left( \mathbf{m}\times \mathbf{h}_{eff} 
+\alpha_{g}\left[ \mathbf{m}\times (\mathbf{m} \times \mathbf{h}_{eff})\right] \right), \nonumber \\ 
\label{LLG}
\end{eqnarray}
where $\textbf{m$\equiv$ m(y,t)}$ is normalized in the units $M_{s}$, which is the saturation magnetization in A/m, the ferromagnetic resonance frequency $\Omega_{F}$ is normalized in the units  $\omega_{J}$, and $\alpha_{G}$ is the Gilbert damping.
\begin{figure*}[t!]
	\center{
	\includegraphics[width=0.7\linewidth]{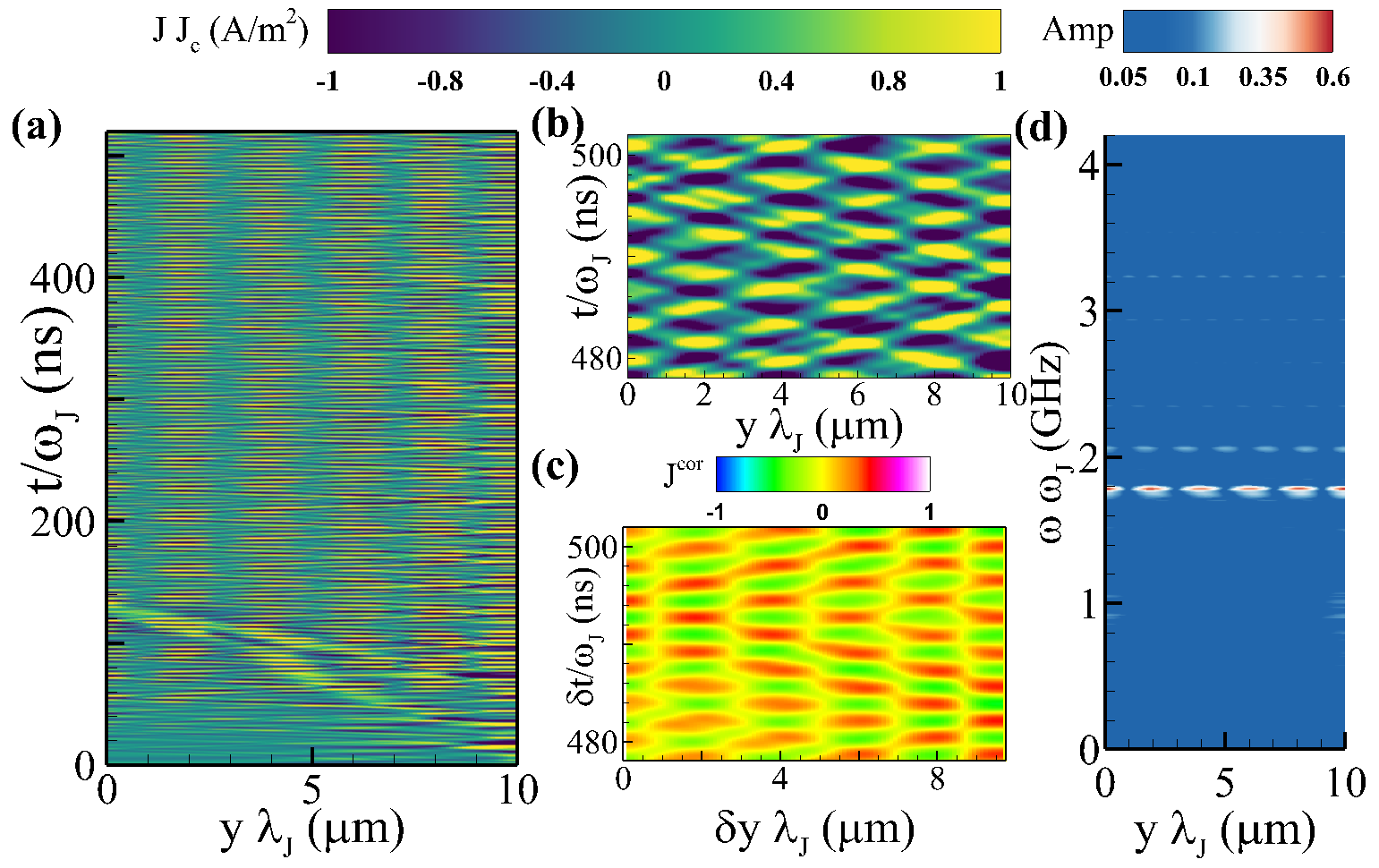}}
	\caption{(a) Spatio-temporal dependence of the in-plane current $J(y,t)$ in the SFS-TI JJ for $c_{ex}=0.05$, $D_{1}=1.1$, and $D_{2}=0.8$.  (b) Magnified view of the part in (a) that demonstrates the STC pattern. (c) The corresponding spatiotemporal averaged current-current correlation function $J^{cor}(\delta y, \delta t)$. (d) The Fast-Fourier transform for the $J(y,t)$ shown in (a). The parameters were fixed to the following values: $\Omega_{F}=1$, $\alpha_{g}=0.05$, $k_{an}=0.5$, $r=0.9$, $G=0.1$, $\beta=0.00024$, $\tilde{d}=0.3$, $\Gamma=0.0762$ and $J_{noise}$ $10^{-4}$. 
    }
	\label{Fig3}
\end{figure*}
The effective field $h_{eff}$ is normalized in the units of $\mu M_{s}$ ($\mu$ is the permeability of the ferromagnet in H/m units), and consists of the fields due to: exchange interaction (ExI) $\textbf{h}_{ex}$, magnetic anisotropy $\textbf{h}_{an}$, Josephson energy $\textbf{h}_{JJ}$ and  DMI $\textbf{h}_{DMI}$ given by~\cite{nashaat2019electrical,golovchanskiy2017ferromagnetic,hong2017magnetic}:
\begin{eqnarray}
\textbf{h}_{ex}  &=& c_{ex} \boldsymbol{\nabla^{2}}m_{i} \hat{\textbf{e}}_{i}, \ \ \ \ \ \ \ \ \  
\textbf{h}_{an}  =  k_{an} m_{i} \hat{\textbf{e}}_{y}, \nonumber \\
\textbf{h}_{JJ} &=& -\Gamma \boldsymbol{\nabla}_{m}[ J_{c}(m_{x}) (1-\cos(\varphi-r m_{y})], \nonumber \\
J_{c}(m_{x})&=& \int_{-\pi/2}^{\pi/2} \cos\theta e^{-\tilde{d}/\cos\theta} \cos\left( r m_{x}  \tan\theta\right) d\theta,\nonumber\nonumber \\
\textbf{h}_{DMI} &=& - \textbf{D}_{1} \times \textbf{m}_{i-1}+ \frac{\textbf{D}_{1}+\textbf{D}_{2}}{2}\times \textbf{m}_{i+1},
\end{eqnarray}
where $c_{ex}$ is the parameter that characterizes ExI, $\Gamma$ is proportional to the ratio of the Josephson and magnetic energy
$\Gamma=G r/J_{c}^{m_{x=0}}$, 
$\theta$ is the angle of the quasiparticle trajectory with respect to the x-axis~\cite{nashaat2019electrical}.
$G=\epsilon_{J}/V_{F}\mu M_{s}^{2}$, $\tilde{d} =2\pi K_{B} T t_{b}/ \hbar v_{F}$ is the dimensionless junction length,  $V_{F}$ is the volume of ferromagnet, $K_{B}$ is the Boltzmann constant, $T$ is the temperature, and $v_{F}$ is the Fermi velocity. 
Due to spin-orbit coupling, the critical current $J_c(m_x)$ is internally modulated by the magnetic moment.
We assume that in the ferromagnet, the DMI constants between $m_{i-1}$ and $m_{i}$ is $D_{1}$, and for $m_{i+1}$ and $m_{i+2}$ is $D_{2}$, where the subscripts $(i-1, i, i+1)$ refer to the space dependence of magnetization~\cite{hong2017magnetic}. 
Thus, the DMI between $m_{i}$ and $m_{i+1}$ is $(D_{1}+D_{2})/2$. 
This means that the DMI constant changes from $D_{1}$ to $D_{2}$ between two magnetic moments~\cite{hong2017magnetic}.

An approximate estimation of the model parameters based on Refs.~\cite{nashaat2019electrical,hong2017magnetic,golovchanskiy2017ferromagnetic,birge2024ferromagnetic} can be done as follow: 
The conventional range of the exchange constant $A_{ex}=10^{-12}-10^{-11}$ J/m, and the DMI can take the range $10^{3}-10^{7}$ J/m$^{3}$ depending on the material. 
Assuming $A_{ex}=5\times10^{-11}$ J/m, DMI constant $D=1\times 10^{4} $ J/m$^{3}$, the saturation magnetization $M_{s}=5\times 10^{4}$ A/m, the anisotropy constant $K_{an}=2$ kJ/m$^{3}$, the London penetration depth $\lambda_{L}=0.5$ $\mu$m, and the Josephson penetration depth $\lambda_{J}= 2.1$ $\mu$m, in this case the exchange length $c_{ex}=l_{ex}^{2}/\lambda_{J}^{2}=0.007$ with $l_{ex}=\sqrt{2 A_{ex}/\mu M_{s}^{2}}$, where $\mu$ is the ferromagnet permeability, $k_{an}=K_{an}/\mu M_{s}^{2}=0.637$ is the normalized anistropy constant, depending on the junction critical current, the coupling constant $G=\epsilon_{J}/(V_{F} \mu M_{s}^{2})=0.063$ for $I_{c}=9\times 10^{-4}$ A, and $D_{1(2)}=3.183$ is the normalized DMI constant with respect to $\mu M_{s}^{2}$. 
For $M=1.9\times10^{4}$ A/m, $A_{ex}=8\times10^{-11}$ J/m, $D=1$ kJ/m$^{3}$, $K_{an}=400$ J/m$^{3}$, $\lambda_{L}=0.1$ $\mu$m, one can estimate $G=0.435$, $k_{an}=0.882$, $D=2.2$, and $c_{ex}=0.016$ for $\lambda_{L}=0.5$ $\mu$m, we have $c_{ex}=0.081$.

The Eqs.~(\ref{sine-Gor_eq}) and (\ref{LLG}) are solved using the Gauss–Legendre quadrature method.
Since the SFS-TI JJ exhibits the splitting of the ferromagnetic easy axis and an unconventional four-fold degenerate ferromagnetic state, the numerical analysis as in Ref.~\cite{nashaat2019electrical} was applied.
The initial conditions for the Josephson phase and magnetic moments are given as $4 \arctan (a_{1}=\cosh( a_{2} x))$ ($a_{1}= 2.5$ and $a_{2}\approx 1$)~\cite{gulevich2020mitmojco}, and 
$m_{x}(y)=m_{z}(y)=0.1$ and $m_{y}(y)=\sqrt{1-m_{x}(y)^{2}-m_{z}(y)^{2}}$, respectively.
During all simulations, we ensure that the total magnetic moment is conserved $\mid\mid m(y,t)\mid\mid =1$.
The in-plane current is analyzed for different values of parameters that characterize ExI and DMI, i.e., $c_{ex}$ and $D_{1,2}$. 

\section{Long $\varphi_0$ Josephson junction on a topological insulator}
\label{SFS_TI JJ}

\begin{figure*}[t!]
	\center{
	\includegraphics[width=0.7\linewidth]{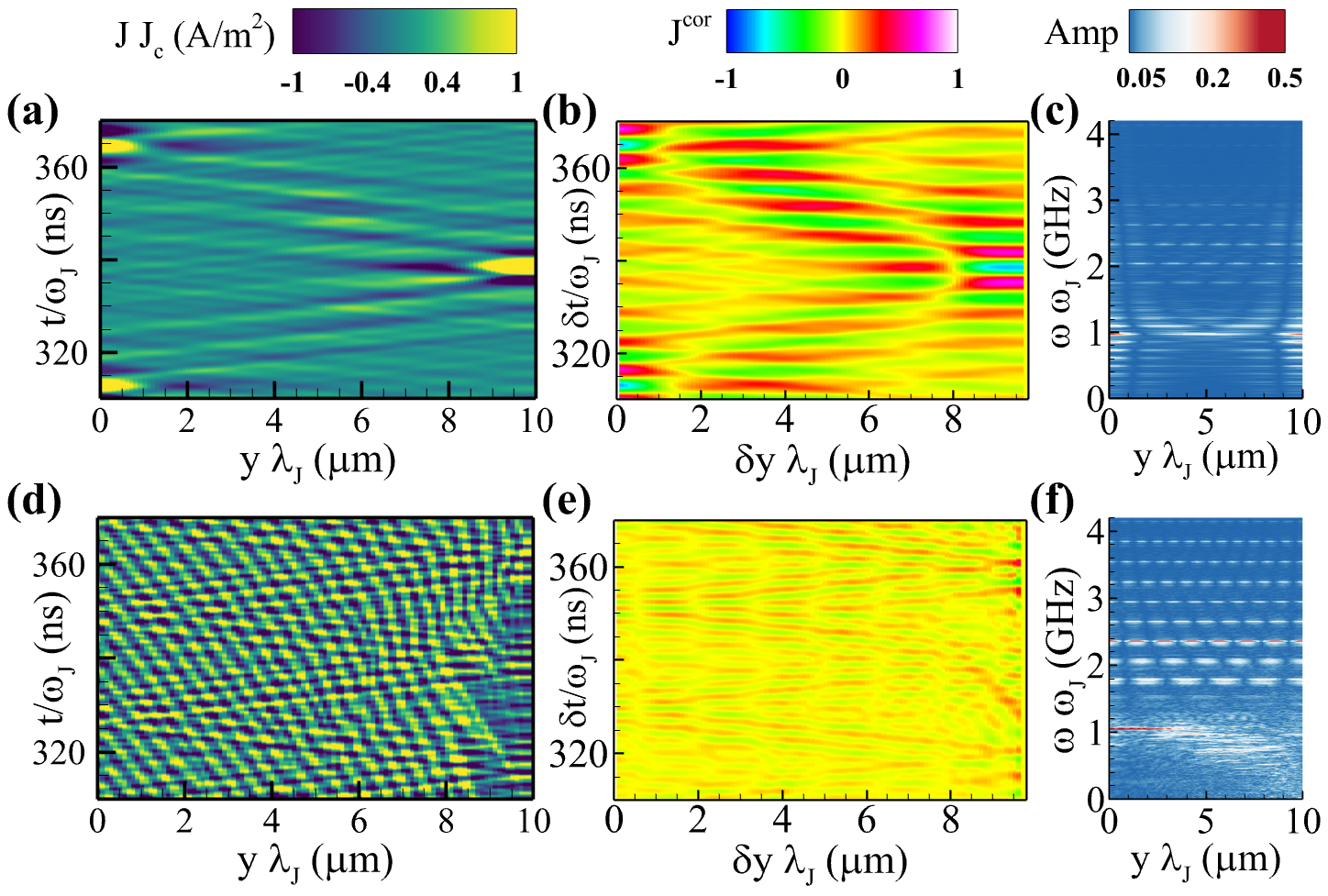}}
	\caption{Spatiotemporal dependence of the in-plane current $J(y,t)$, the current-current correlation function $J^{cor}(\delta y,\delta t)$ and the corresponding FFT analysis for DMI$=0$ and ExI$\neq 0$  (qualitatively the same for ExI$=$DMI$=0$) in (a)-(c), and DMI$\neq0$ and ExI$=0$ in (d-f), respectively.
    The rest of the parameters are the same as in Fig.~\ref{Fig3}.
 }
	\label{Fig4}
\end{figure*}

To investigate the emergence of spacetime crystalline order in the system presented in Fig.~\ref{Fig1}, we follow the procedure similar to one used in Ref.~\cite{kleiner2021space}, where a classical discrete space-time crystal was observed under external periodic modulation of the critical current in systems of intrinsic conventional Josephson junctions.
For different parameter settings, we investigate the spatiotemporal dependence of the in-plane current and the current-current correlation function, perform the FFT-analysis, and examine the robustness of the observed STC pattern. 
In addition, the behavior of the magnetic subsystem is also investigated.

\subsection{The occurrence of STC order}
\label{STC}

In Fig.~\ref{Fig3}(a) and (b), the spatiotemporal dependence of the in-plane current $J(y,t)$ in the SFS-TI JJ is presented. As Fig.~\ref{Fig3}(a) demonstrates a propagating standing wave along the junction, the magnified view shown in Fig.~\ref{Fig3}(b) reveals the STC order and breaking of the continuous translational symmetry in both space and time. 
This symmetry is classified as $C_{2}m_{x}m_{t}$ with rhombohedral unit cells~\cite{xu2018space}. 
This pattern is similar to the case demonstrated in Ref.~\cite{kleiner2021space} for a long SIS Josephson junction under the external modulation of the critical current. 

The way to distinguish the STC order from other collective oscillations is to analyze the correlation function and the fast Fourier transform for the in-plane current and magnetic moments.
This can provide a direct signature of the STC pattern.
The manifestation of the STC pattern
requires that the long-range correlation function across the junction must be large enough. 
The spatiotemporal averaged current-current correlation function  $J^{cor}(\delta y,\delta t)$ in this case is defined as~\cite{kleiner2021space}:
\begin{equation}
J^{cor}(\delta y,\delta t) = \frac{\langle J(y,t) J(y+\delta y, t+\delta t)\rangle_{y,t}}{\sqrt{\langle J^2 (y,t)\rangle_{y,t}\langle J^2 (y+\delta y,t+\delta t)\rangle_{y,t}}},
\end{equation}
where the averaging is taken over both space and time. 
In Fig. \ref{Fig3}(c), we show that $J^{cor}(\delta y,\delta t)$ also exhibits the STC pattern.

To determine the oscillation frequency of the in-plane current in Fig.~\ref{Fig3}(b), we analyze the FFT for the current $J(y,t)$.
In Fig.~\ref{Fig3}(d), the results of the FFT analysis show that the system tends to pick favorable modes within the band around $2\Omega_{F}$ and oscillates at its own frequency $\omega=1.8 \Omega _F$.

The presence of both DMI and ExI represents the key requirement, since their interplay leads to the appearance of the STC order.
This is indicated in Fig.~\ref{Fig4}, where the spatiotemporal dependence of the in-plane current $J(y,t)$ is presented in the case when DMI or ExI is absent.
The STC order is not observed in either of these two cases.
While the result in Fig.~\ref{Fig4}(a) reveals traveling waves, the correlation function in (b) clearly shows the absence of any STC order.
The FFT in Fig.~\ref{Fig4}(c) shows frequency lines with a broadened band around the FMR frequency $\omega =0.97\Omega_{F}$.
We stress that the resonance frequency in an SFS junction is shifted from the FMR due to the coupling between the Josephson phase and magnetization, spin-orbit coupling, and Gilbert damping~\cite{shukrinov2019ferromagnetic,shukrinov2021anomalous}.
The presented results for the case with DMI$=0$, and ExI $\neq0$ are qualitatively the same as if both interactions were absent.

In the case when the DMI is taken into account while ExI is absent, as presented in Fig.~\ref{Fig4}(d)-~\ref{Fig4}(f), we can see some "zebra-like" patterns in Fig.~\ref{Fig4}(d).
As in the previous case in Fig.~\ref{Fig4}(b), the correlation function in Fig.~\ref{Fig4}(e) shows the absence of any STC order.
The presence of DMI excites the higher modes in the system, and the FFT result in Figure~\ref{Fig4}(e) demonstrates higher harmonics and subharmonics of the FMR frequency with a maximum amplitude for $\omega=0.97 \Omega_{F}$, and $\omega=2.36\Omega_{F}$.

\begin{figure}[h!]
	\centering
	\includegraphics[width=\linewidth]{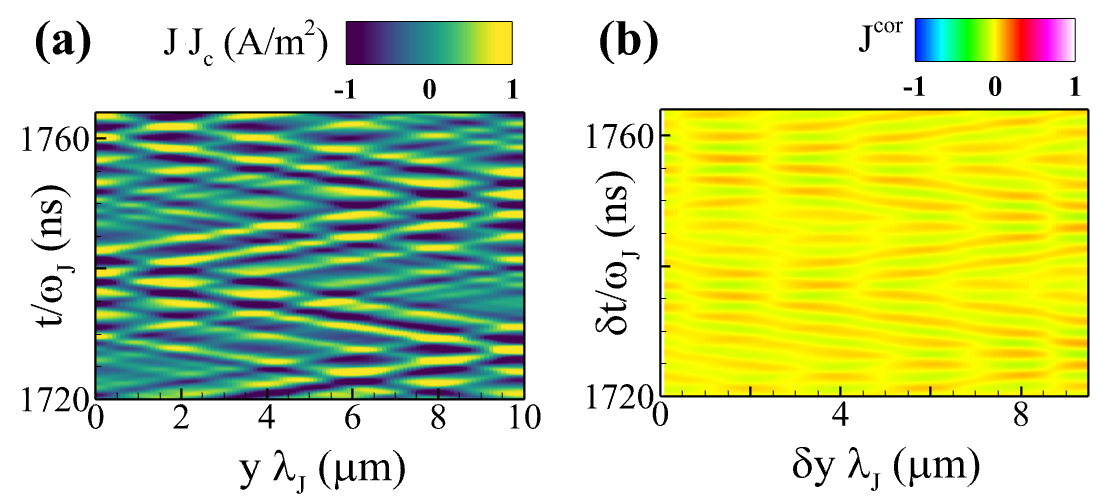}
	\caption{ (a) Magnified view of the spatiotemporal dependence of the in-plane current $J(y,t)$. (b) The corresponding averaged current-current correlation function $J^{cor}(\delta y,\delta t)$.  All panels are created for the same parameters as in Fig.~\ref{Fig3}, but in the extended time domain.}
	\label{Fig5}
\end{figure}

Other important properties of the STC order are its robustness and lifetime. 
Fig.~\ref{Fig5} demonstrates the STC pattern shown in Fig.~\ref{Fig3} but in a larger time domain.
From the magnified view of the current diagram $J(y,t)$ shown in Fig.~\ref{Fig5}(a) and the corresponding averaged current-current correlation function in Fig.~\ref{Fig5}(b), we can see that in a longer time domain, the system still exhibits STC order.
In our simulations, we extended the time up to 60 times the original simulation time, despite the reduction in the magnitude of the current J(y,t) and $J^{corr}(y,t)$, the spatiotemporal order remains intact. We also increased the length of the junction 10 times, and the STC occurred depending on the parameters of the junction.
 
\subsection{The magnetic subsystem}
\label{M}

\begin{figure}[h!]
\centering
\includegraphics[width=0.9\linewidth]{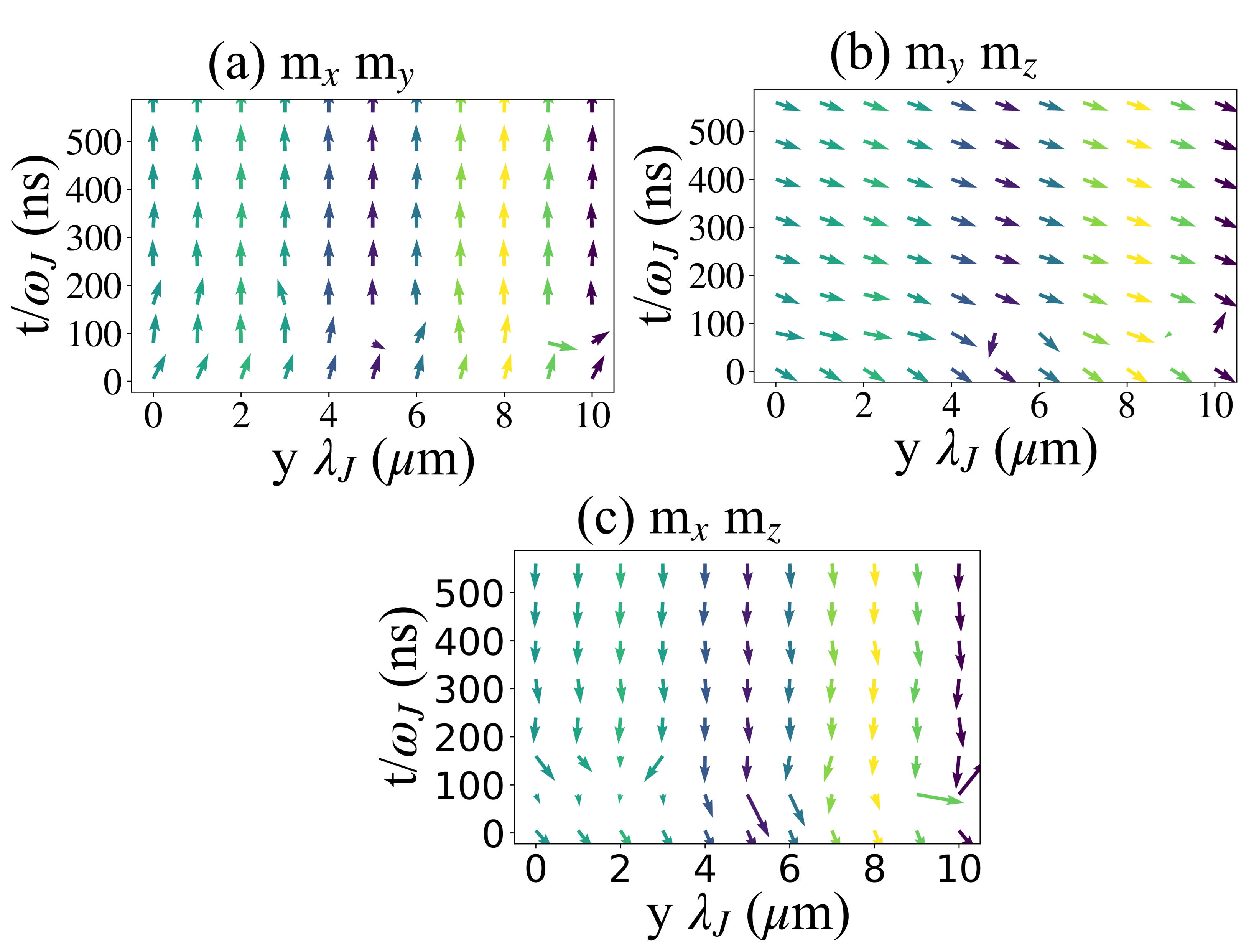}
\caption{[(a)-(c)] In-plane magnetization vector fields for the case shown in Fig. \ref{Fig3}}
    \label{Fig6}
\end{figure}

So far, we have been focused only on the spatiotemporal dependence of the in-plane current; meanwhile, the behavior of the magnetization components has not been considered.
In Fig.~\ref{Fig6}, we present the screenshots for the magnetization vector fields for the same parameters as in Fig.~\ref{Fig3}.
As we see, they all relax in the same direction.

\begin{figure}[h!]
	\centering
	\includegraphics[width=\linewidth]{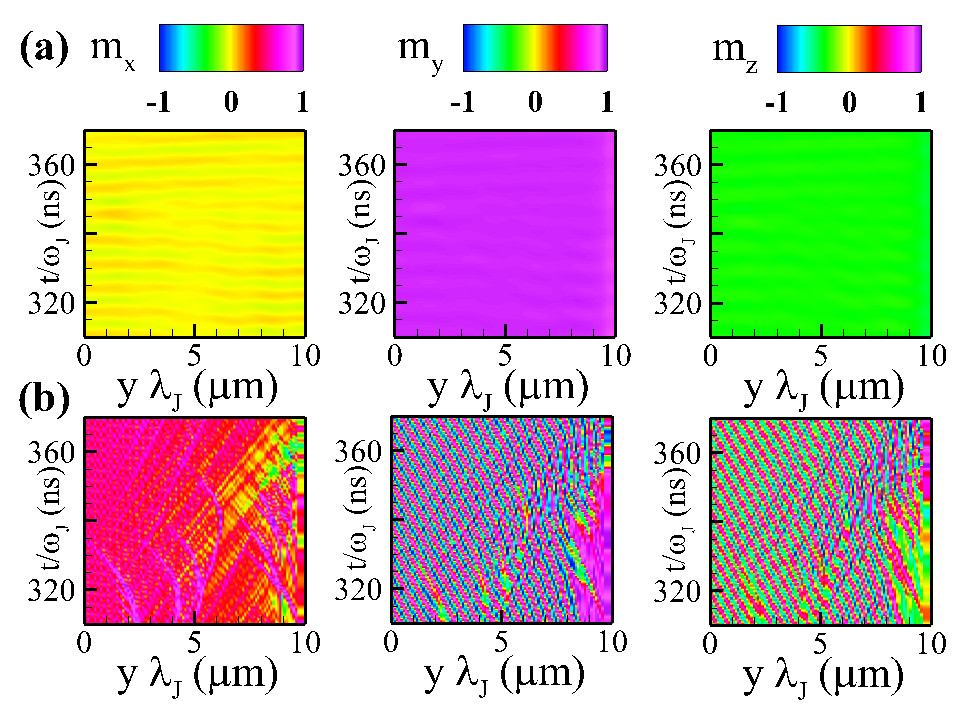}
	\caption{ Magnified view of the spatiotemporal dependence of the magnetic components in $x,y,$ and $z$ direction: (a) ExI$\ne0$ and DMI$\ne0$ and (b) ExI$=0$ and DMI$\neq0$. All panels are done with the same parameters as in Fig.~\ref{Fig3}.}
	\label{Fig7}
\end{figure}

Furthermore, we did not observe the STC pattern in the spatiotemporal dependence of the magnetic components. 
This is illustrated in Fig.~\ref{Fig7}, where 
$m_{x,y,z}(y,t)$ components are plotted in two cases. 
The first, in Fig.~\ref{Fig7}(a), corresponds to the STC order in Fig.~\ref{Fig3}.
As we can see, while the current demonstrates the STC order in Fig.~\ref{Fig3}(b), there is no sign of the STC pattern in Fig.~\ref{Fig7}(a).
In fact, the spatiotemporal dependence of the magnetic components shows no signature of phase coherence or collective oscillations with the in-plane current.

The second case in Fig.~\ref{Fig7}(b)  illustrates an example of the collective oscillations and corresponds to the spatiotemporal dependence of the in-plane current in Fig.~\ref{Fig4}(d) for ExI$=0$ and DMI$\ne0$.
We can see that the patterns in Fig.~\ref{Fig7}(b) are very similar to those in Fig.~\ref{Fig4}(d). 
The FFT analysis for $m_{y(z)}(y,t)$ (not shown here) demonstrates one frequency band around $\omega=\omega_{F}$, as the one in Fig.~\ref{Fig4}(f).
This shows that the two subsystems are synchronized and oscillate collectively. 
Hence, the results in Fig.~\ref{Fig7} together with those from Subsec.~\ref{STC} provide a clear distinction between the STC order and the collective oscillations.

\begin{figure*}[t!]
	\center{
	\includegraphics[width=0.6\linewidth]{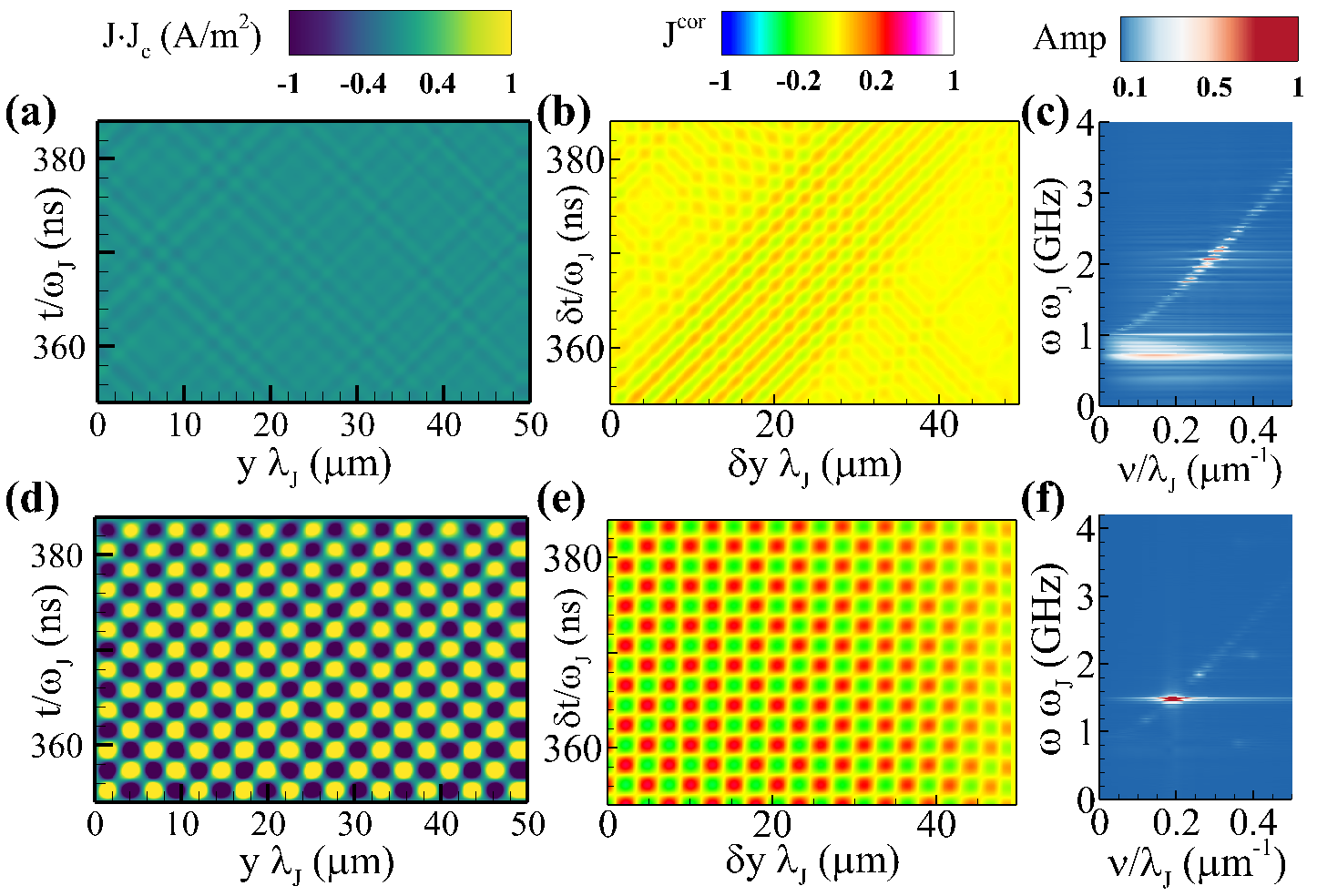}}
	\caption{Spatiotemporal dependence of the in-plane current $J(y,t)$, the averaged current-current correlation function $J^{cor}(\delta y,\delta t)$ and the corresponding 2D FFT analysis for a long $\varphi _0$ Josephson junction in (a-c) and the case when the external modulation with the amplitude $A=0.15$ and the frequency $\omega_{mod} =3$ ia applied.
    The rest of the parameters are the same as in Fig.~\ref{Fig3}.
 }
	\label{Fig8}
\end{figure*}

\section{Long $\varphi _0$ Josephson junction}
\label{SFS JJ}

If we consider only a long $\varphi _0$ Josephson junction and the limit where the critical current remains constant and unaffected by the dynamics of the magnetic moment~\cite{abdelmoneim2022locking,guarcello,guarcello2023switching,bobkov2024controllable}, then no STC order can appear, as can be seen from $J(y,t)$, and $J^{cor}(\delta y,\delta t)$, in Fig.~\ref{Fig8}(a) and \ref{Fig8}(b), respectively.

However, if we apply an external parametric modulation as was done in Ref.~\cite{kleiner2021space}, then the results would change completely. 
Following the same method, the critical current is parametrically modulated from the external source with frequency $\omega_{mod}$ normalized to $\omega_{J}$ and amplitude $A_{mod}$ normalized to $J_{c}$.
The supercurrent in this case reads as $J_{s}= J_{mod} \sin(\varphi-r m_{y})$, where $J_{mod}$ is normalized to $J_{c}$ and given by~\cite{kleiner2021space}:
\begin{eqnarray}
J_{mod}=\frac{1+ A_{mod}\cos(\omega_{mod} t)}{1+A_{mod}}.
\end{eqnarray}
In Fig.~\ref{Fig8}(d) and (e), we have a typical example of a classical discrete STC order (for $A_{mod}=0.15$ and $\omega_{mod}=3$), as the one observed in  Ref.~\cite{kleiner2021space} that, as Fig.~\ref{Fig8}(f) shows, oscillates at half the modulation frequency.
In this case, to make a comparison with the results in Ref.~\cite{kleiner2021space}, we have performed the corresponding 2D FFT analysis. 

If we compare our results with the results in Ref.~\cite{kleiner2021space}, then we can see that our crystalline pattern has sharper boundaries.
In Ref.~\cite{kleiner2021space}, they had to extend their study from the 1D single-layer long junction to the 2D  and 3D multilayer junctions stacks to get a more pronounced and stable STC order. 
Meanwhile, in our case, even a simple long $\varphi_0$ Josephson junction under the same type of modulation exhibits a well-defined STC pattern.
Our results in Fig.~\ref{Fig8}(a) and (b) show that for the realization of a classical discrete time crystal, a hybrid JJ with a ferromagnetic interface might be a better choice than a conventional junction used in Ref.~\cite{kleiner2021space}.

\section{Discussion}
\label{Dis}

For an STC order to appear in a Josephson junction system, the modulation of the critical current is mandatory~\cite{kleiner2021space}, and here it was achieved internally.
Our system consists of coupled ferromagnetic and superconducting subsystems. 
In the ferromagnetic one, the exchange interaction tends to align the magnetic moments in the same direction.
This affects Cooper pairs traveling through a ferromagnetic layer, leading to oscillations in the pair amplitude and consequently to the suppression and oscillations of the supercurrent~\cite{buzdin2005proximity}.
On the other hand, DMI favors a noncollinear order of magnetic moments, which changes the distribution of their polarization across the junction. The interplay between ExI and DMI modulates the precession of the magnetic moment inside the ferromagnet. 
This further influences the superconducting subsystem through the modulation of the critical current and the additional phase shift in the current-phase relation.
This internal modulation of the critical current leads to the appearance of STC order in Fig.~\ref{Fig3}(b) and (c). 

To get a deeper insight into the physics behind the observed STC order, we also performed an ablation investigation.
The obtained results show that, when ExI and DMI are excluded in Fig.~\ref{Fig4}, or in the limit when the critical current is constant in Fig.\ref{Fig8}(a-c), the STC pattern disappears in the spatiotemporal dependence of $J(y,t)$ and $J^{cor}(\delta y, \delta t)$.
In addition, the FFT analysis in Fig.~\ref{Fig4}(c) and \ref{Fig4}(f) and Fig.~\ref{Fig8}(c) shows that the system oscillates at the FMR frequency and its harmonics and subharmonics.
However, this is not the case when the STC order appears as we see in Fig.~\ref{Fig3}(d), where the bands around $\omega _F$ completely disappear.

A clear difference between the STC order and the collective oscillations is demonstrated in the behavior of the magnetic subsystem in Sec.\ref{M}.
In the case of the STC order in Fig.~\ref{Fig3}, the corresponding behavior of the magnetic moment in Fig.~\ref{Fig7}(a) shows that the magnetic subsystem does not exhibit STC behavior, despite being coupled to the superconducting one and the Josephson current affecting the magnetic moment through the Josephson effective field.
This is completely different from the case of collective oscillations illustrated in Fig.~\ref{Fig4}(d) and Fig.~\ref{Fig7}(b), where both systems, being coupled, oscillate in synchrony.

In the limit when the critical current is not modulated by the magnetic moment,  the STC order can be achieved if and only if an external periodic modulation is applied.
The results in Fig.~\ref{Fig8}(d) and \ref{Fig8}(e) represent an example of a classical discrete STC pattern that oscillates at half the modulation frequency, and it is consistent with the findings of Ref.~\cite{kleiner2021space}.
If we compare our results in Fig.~\ref{Fig3} and Fig.~\ref{Fig8}(d-f) with the previous studies in Ref.~\cite{kleiner2021space}, then we see that all these systems behave in a similar fashion: Breaking of the time translation symmetry is characterized by the appearance of the STC pattern in the spatiotemporal dependence of the in-plane current and the corresponding correlation functions that oscillate at one particular frequency different from the modulation frequency.
However, the STC pattern in Fig.~\ref{Fig3} has a different physical origin: The critical current is not externally modulated as in the case in Fig.~\ref{Fig8}(d)-~\ref{Fig8}(f) or in Ref.~\cite{kleiner2021space} but internally by the magnetic moment~\cite{nashaat2019electrical}.

In general, discrete time crystals occur due to breaking of discrete time translation symmetry in closed nonequilibrium systems as a subharmonic response to the external periodic modulation, i.e., they oscillate at half the modulation frequency~\cite{machado2023absolutely,yi2024theory, russomanno2023spatiotemporally}. 
Continuous-time crystals, on the other hand, come from breaking the continuous time translation symmetry in open systems without external periodic input and are characterized by spontaneous oscillations~\cite{zaletel2023colloquium,kongkhambut2024observation}.
In our case, the intrinsic STC order in Fig.~\ref{Fig3} has been realized in an open system without external periodic modulation.
However, its periodicity is not arbitrary, i.e., spontaneously chosen from a continuous symmetry~\cite{kessler2021observation,kongkhambut2024observation,chen2023realization,yang2025emergent}. 
Due to the presence of FMR, Josephson junctions with a ferromagnetic interface possess an internal clock, and consequently, in our case, the periodicity of the STC order is tied to an internal mode determined by the FMR frequency. 

The strict definition of a time crystal requires that its existence be proven in the thermodynamic limit, i.e., when the degrees of
freedom, the system size, and time all go to infinity.
While discrete time crystals can achieve the thermodynamic limit and are often considered as "true" time crystals, continuous ones remain more controversial since they manifest as nonequilibrium dynamical phases with long-range temporal order in open nonequilibrium systems.
However, as the definition of time crystals evolved from the original Wilczek idea, the same was also true for the thermodynamic limit requirement.
The thermodynamic limit, which involves infinitely large systems, has not always been addressed or proven in the works on time crystals~\cite{kleiner2021space,yao2020classical,kyprianidis2021observation}.
Instead, the findings were specific to the experimental setup and finite system size.
In Ref.~\cite{kleiner2021space}, the authors note that increasing the system size can lead to strong mode competition, which can result in chaotic behavior.
They further speculate that in the thermodynamic limit, only modes with a spatial period on the order of the Josephson length may survive.
In Ref.~\cite{kyprianidis2021observation}, for example, the study demonstrates the robust prethermal discrete time crystal in a controlled experimental setup using a 25-ion chain in a trapped-ion quantum simulator.
In the current experimental setups, no time crystal can last forever.
Their lifetime is always finite, even though they can exhibit remarkably long-lived temporal order.
The intrinsic STC order we observed in our work was stable over a wide range of parameters.
The results presented here also indicate that the SFS Josephson junction under external modulation might be a better choice than the conventional Josephson junctions used in Ref.~\cite{kleiner2021space}, but to confirm this, more investigation on the effect of spin-orbit coupling on the STC order is needed.
How far we can go by increasing time and the system size, as well as the effect of noise and other parameters, will be studied in the future.

\section{Experimental validation}
\label{Exp}

\begin{figure}[h]
\centering
\includegraphics[width=\linewidth]{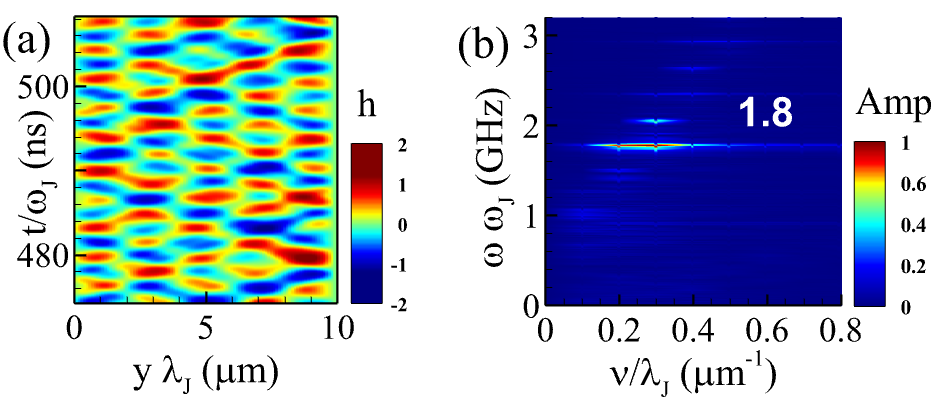}
\caption{(a) Prediction for the in-plane spatiotemporal magnetic field (h). (b) Spatiotemporal fast Fourier transform. Both panels are obtained for the case shown in Fig.~\ref{Fig3}.}
    \label{Fig9}
\end{figure}

Experimental detection of the STC order in hybrid JJs involves multiple challenges, from providing junctions with the right quality and geometry to solving environmental issues and finding proper measurement techniques. 
Since no experiments have been carried out so far on the detection of STC order in Josephson junction systems, finding the right way to detect them, i.e., to measure the spatiotemporal dependence of current $J(y,t)$, is of utmost importance.
Recently, an experimental platform was introduced for visualizing at a nano-scale the supercurrent flow in JJ~\cite{chen2024current} and probing the Abrikosov vortices in niobium~\cite{hou2024probing}.
The authors in both works used the scanning magnetometry device with the nitrogen-vacancy centers, which can be optically manipulated and are magnetically sensitive. 
If this device is placed near a Josephson junction, then one can detect the changes in the magnetic field strength caused by a supercurrent flow and give real-time imaging of the spatiotemporal distribution of the supercurrent throughout the junction.
Thus, this approach can pave the way for the detection of the STC order in the JJ, which is reflected in the intrinsic magnetic field of the junction.
This is demonstrated in Fig.~\ref{Fig9} where the intrinsic magnetic field in the Josephson junction is determined by the coordinate derivative of the phase difference~\cite{rahmonov2014parametric} in units $J_{c}\lambda_{J}$.
The result in Fig.~\ref{Fig9}(a) indicates that the STC pattern is manifested in the spatiotemporal dependence in the magnetic field of the JJ.
In experiments with current biased junctions, the total voltage across the junction is changing. 
These voltage variations are related to the local time average for the current and can be mapped via scanning laser microscopy~\cite{kleiner2021space}.
Also, to show that the in-plane current oscillates at the frequency $1.8\Omega_F$ (see Fig.~\ref{Fig4}), we present in Fig.~\ref{Fig9}(b) the expected spectrum for the fast Fourier-transform infrared spectroscopy for the temporal dependence of the current, which
demonstrates a strong frequency band around $1.8\Omega_{F}$. 

\section{Conclusion}
\label{Con}

In this work, we have shown that a long $\varphi_0$ Josephson junction placed on a topological insulator can exhibit an intrinsic space-time crystalline order without any external periodic input.
Due to the coupling between the magnetic moment and Josephson phase, the presence of the exchange and Dzyaloshinskii–Moriya interactions in the ferromagnetic layer modulates the critical current internally.
This leads to the appearance of the STC order in the spatiotemporal dependence of the in-plane current and the averaged current-current correlation function that oscillates at almost twice the FMR frequency.
In the case of a long $\varphi_0$ JJ, in the limit when the Josephson critical current is unaffected by the magnetic moment, the STC pattern disappears.
In such a case, an STC order can be realized only under an external parametric periodic modulation of the critical current, in which case we obtain a typical example of a classical discrete time crystal, as the one previously observed in a stack of long Josephson junctions~\cite{kleiner2021space}.
In addition, we also investigate the possibilities for experimental detection of STC order and show that a recently developed magnetometry device that visualizes a supercurrent flow at the nanoscale can be used to detect the STC pattern in the current.

Though this work was dedicated to one particular type of hybrid Josephson junction, we need to stress that the above conclusions can be generally applied to other SFS heterostructures, where the critical current is internally modulated by the magnetic moment.
One such example is the junctions with the multilayer ferromagnetic interface and broken structural inversion symmetry.
These studies will be presented in the future.

The realization of time crystals in hybrid Josephson junctions could bring significant advances in physics across multiple fields~\cite{zaletel2023colloquium}. 
The merging of nonequilibrium quantum states with superconducting and magnetic systems could allow the exploration of new phases of quantum matter and topological quantum states~\cite{abanin2019colloquium,hannaford2022decade,xu2023boundary}.
As inherently low-energy systems, by increasing stability, efficiency, and precision, time crystals could push boundaries in the engineering of classical and quantum devices from quantum computing~\cite{kockum2019quantum} and spintronics~\cite{linder2015superconducting,incorvia2024spintronics} to metrology~\cite{choi2017quantum}.

\section*{Acknowledgments}
The authors thank I. R. Rahmonov and K. Kulikov for the fruitful discussion. M. Nashaat and Yu. M. Shukrinov acknowledge Cairo University (Egypt), BLTP (Russia), within the Cooperation Agreement between ASRT, Egypt and JINR, Russian Federation.  M. Nashaat acknowledges the financial support of the Russian Science Foundation in the framework of Project No. 22-71-10022 of the Russian Scientific Fund. J. T. acknowledge the support from the Ministry of Education, Science, and Technological Development of the Republic of Serbia, Grant No. 451-03-1F36/2025-03/ 200017 (“Vinča” Institute of Nuclear Sciences, University of Belgrade) and the Projects within the Cooperation Agreement between the JINR, Dubna, Russian Federation and the Republic of Serbia (P02 and P12). Special thanks to the Laboratory of Information Technologies of the Joint Institute for Nuclear Research for the opportunity to use the computational resources of the HybriLIT, BLTP (JINR), and the Bibliotheca Alexandrina’s (Egypt) for the High-Performance Computing infrastructure.

\bibliography{SFSTIJJSTCbib}
\end{document}